\RequirePackage{lineno}
\documentclass[aps,twocolumn,superscriptaddress,preprintnumbers,amsmath,amssymb,floatfix,nofootinbib]{revtex4}

\usepackage{graphicx} 
\usepackage{dcolumn}  
\usepackage{rotating} 
\usepackage{lscape}

\usepackage{bm}
\usepackage{url} 
\usepackage{amsmath} 
\usepackage{amssymb} 

\usepackage{verbatim}
\usepackage{multirow}
\usepackage{subfigure}

\graphicspath{{ps}}

\begin{document}

\title{Search for the Sagittarius Tidal Stream of Axion Dark Matter around 4.55 $\mu$eV}
\author{Andrew K. Yi}\affiliation{Dept.\ of Physics, Korea Advanced Institute of Science and Technology, Daejeon 34141, Republic of Korea}\affiliation{Center for Axion and Precision Physics Research, Institute for Basic Science, Daejeon 34051, Republic of Korea}
\author{Saebyeok Ahn}\affiliation{Dept.\ of Physics, Korea Advanced Institute of Science and Technology, Daejeon 34141, Republic of Korea}\affiliation{Center for Axion and Precision Physics Research, Institute for Basic Science, Daejeon 34051, Republic of Korea}
\author{\c{C}a\u{g}lar Kutlu}\affiliation{Dept.\ of Physics, Korea Advanced Institute of Science and Technology, Daejeon 34141, Republic of Korea}\affiliation{Center for Axion and Precision Physics Research, Institute for Basic Science, Daejeon 34051, Republic of Korea}
\author{JinMyeong Kim}\affiliation{Dept.\ of Physics, Korea Advanced Institute of Science and Technology, Daejeon 34141, Republic of Korea}\affiliation{Center for Axion and Precision Physics Research, Institute for Basic Science, Daejeon 34051, Republic of Korea}
\author{Byeong Rok Ko}\email[Corresponding author~:~]{brko@ibs.re.kr}\affiliation{Center for Axion and Precision Physics Research, Institute for Basic Science, Daejeon 34051, Republic of Korea}
\author{Boris I. Ivanov}\affiliation{Center for Axion and Precision Physics Research, Institute for Basic Science, Daejeon 34051, Republic of Korea}
\author{HeeSu Byun}\affiliation{Center for Axion and Precision Physics Research, Institute for Basic Science, Daejeon 34051, Republic of Korea}
\author{Arjan F. van Loo}\affiliation{RIKEN Center for Quantum Computing (RQC), Wako, Saitama 351-0198, Japan}\affiliation{Dept.\ of Applied Physics, Graduate School of Engineering, The University of Tokyo, Bunkyo-ku, Tokyo 113-8656, Japan}
\author{SeongTae Park}\affiliation{Center for Axion and Precision Physics Research, Institute for Basic Science, Daejeon 34051, Republic of Korea}
\author{Junu Jeong}\affiliation{Center for Axion and Precision Physics Research, Institute for Basic Science, Daejeon 34051, Republic of Korea}
\author{Ohjoon Kwon}\affiliation{Center for Axion and Precision Physics Research, Institute for Basic Science, Daejeon 34051, Republic of Korea}

\author{Yasunobu Nakamura}\affiliation{RIKEN Center for Quantum Computing (RQC), Wako, Saitama 351-0198, Japan}\affiliation{Dept.\ of Applied Physics, Graduate School of Engineering, The University of Tokyo, Bunkyo-ku, Tokyo 113-8656, Japan}
\author{Sergey V. Uchaikin}\affiliation{Center for Axion and Precision Physics Research, Institute for Basic Science, Daejeon 34051, Republic of Korea}
\author{Jihoon Choi}\altaffiliation[]{Now at Korea Astronomy and Space Science Institute, Daejeon 34055, Republic of Korea}\affiliation{Center for Axion and Precision Physics Research, Institute for Basic Science, Daejeon 34051, Republic of Korea}
\author{Soohyung Lee}\affiliation{Center for Axion and Precision Physics Research, Institute for Basic Science, Daejeon 34051, Republic of Korea}
\author{MyeongJae Lee}\altaffiliation[]{Now at Dept.\ of Physics, Sungkyunkwan University, Suwon 16419, Republic of Korea}\affiliation{Center for Axion and Precision Physics Research, Institute for Basic Science, Daejeon 34051, Republic of Korea}
\author{Yun Chang Shin}\affiliation{Center for Axion and Precision Physics Research, Institute for Basic Science, Daejeon 34051, Republic of Korea}
\author{Jinsu Kim}\affiliation{Dept.\ of Physics, Korea Advanced Institute of Science and Technology, Daejeon 34141, Republic of Korea}\affiliation{Center for Axion and Precision Physics Research, Institute for Basic Science, Daejeon 34051, Republic of Korea}
\author{Doyu Lee}\altaffiliation[]{Now at Samsung Electronics, Gyeonggi-do 16677, Republic of Korea}\affiliation{Center for Axion and Precision Physics Research, Institute for Basic Science, Daejeon 34051, Republic of Korea}
\author{Danho Ahn}\affiliation{Dept.\ of Physics, Korea Advanced Institute of Science and Technology, Daejeon 34141, Republic of Korea}\affiliation{Center for Axion and Precision Physics Research, Institute for Basic Science, Daejeon 34051, Republic of Korea}
\author{SungJae Bae}\affiliation{Dept.\ of Physics, Korea Advanced Institute of Science and Technology, Daejeon 34141, Republic of Korea}\affiliation{Center for Axion and Precision Physics Research, Institute for Basic Science, Daejeon 34051, Republic of Korea}
\author{Jiwon Lee}\affiliation{Dept.\ of Physics, Korea Advanced Institute of Science and Technology, Daejeon 34141, Republic of Korea}\affiliation{Center for Axion and Precision Physics Research, Institute for Basic Science, Daejeon 34051, Republic of Korea}
\author{Younggeun Kim}\affiliation{Center for Axion and Precision Physics Research, Institute for Basic Science, Daejeon 34051, Republic of Korea}
\author{Violeta Gkika}\affiliation{Center for Axion and Precision Physics Research, Institute for Basic Science, Daejeon 34051, Republic of Korea}
\author{Ki Woong Lee}\affiliation{Center for Axion and Precision Physics Research, Institute for Basic Science, Daejeon 34051, Republic of Korea}
\author{Seonjeong Oh}\affiliation{Center for Axion and Precision Physics Research, Institute for Basic Science, Daejeon 34051, Republic of Korea}
\author{Taehyeon Seong}\affiliation{Center for Axion and Precision Physics Research, Institute for Basic Science, Daejeon 34051, Republic of Korea}
\author{DongMin Kim}\affiliation{Center for Axion and Precision Physics Research, Institute for Basic Science, Daejeon 34051, Republic of Korea}
\author{Woohyun Chung}\affiliation{Center for Axion and Precision Physics Research, Institute for Basic Science, Daejeon 34051, Republic of Korea}
\author{Andrei Matlashov}\affiliation{Center for Axion and Precision Physics Research, Institute for Basic Science, Daejeon 34051, Republic of Korea}
\author{SungWoo Youn}\affiliation{Center for Axion and Precision Physics Research, Institute for Basic Science, Daejeon 34051, Republic of Korea}
\author{Yannis K. Semertzidis}\affiliation{Center for Axion and Precision Physics Research, Institute for Basic Science, Daejeon 34051, Republic of Korea}\affiliation{Dept.\ of Physics, Korea Advanced Institute of Science and Technology, Daejeon 34141, Republic of Korea}

\begin{abstract}
  We report the first search for the Sagittarius tidal stream of axion
  dark matter around 4.55 $\mu$eV using CAPP-12TB haloscope data
  acquired in March of 2022.  
  Our result excluded the Sagittarius tidal stream of
  Dine-Fischler-Srednicki-Zhitnitskii and
  Kim-Shifman-Vainshtein-Zakharov axion dark matter densities of
  $\rho_a\gtrsim0.184$ and $\gtrsim0.025$~GeV/cm$^{3}$, respectively,
  over a mass range from 4.51 to 4.59~$\mu$eV at a 90\% confidence
  level.
\end{abstract}


\maketitle

{\renewcommand{\thefootnote}{\fnsymbol{footnote}}}
\setcounter{footnote}{0}

Approximately 85\% of matter in our Universe
consists of cold dark matter (CDM) according to the standard model of
big bang cosmology and precision cosmological
measurements~\cite{PLANCK}.
Despite the strong evidence of the existence of dark
matter~\cite{CDM-EVIDENCE}, its nature remains unknown to date and
falls into a category that is beyond the standard model of particle
physics (SM).
One of the strongest CDM candidates is the axion~\cite{AXION}, which
results from the breakdown of a new form of global symmetry introduced
by Peccei and Quinn~\cite{PQ} to solve the strong $CP$ problem in the
SM~\cite{strongCP}. The axion is predicted to be massive, abundant,
nonrelativistic, and interacts very weakly with the SM~\cite{CDM_LOW}.

A direct axion detection method by Sikivie~\cite{sikivie}, also known
as the axion haloscope, uses the axion-photon coupling
$g_{a\gamma\gamma}=\frac{\alpha g_{\gamma}}{\pi f_a}$, where $\alpha$
is the fine structure constant, $g_\gamma$ is a model-dependent
coupling constant, and $f_a$ is the axion decay constant.
The Kim-Shifman-Vainshtein-Zakharov (KSVZ) model~\cite{KSVZ} and the
Dine-Fischler-Srednicki-Zhitnitskii (DFSZ) model~\cite{DFSZ} predict
the $g_{a\gamma\gamma}$ with $g_\gamma=-0.97$ and 0.36, respectively.
Thanks to the resonant conversion of axions to photons in a microwave
cavity, the axion haloscope provides the most sensitive axion dark
matter search via $g_{a\gamma\gamma}$ in the microwave region.

\begin{figure}
  \centering
  \includegraphics[width=0.42\textwidth]{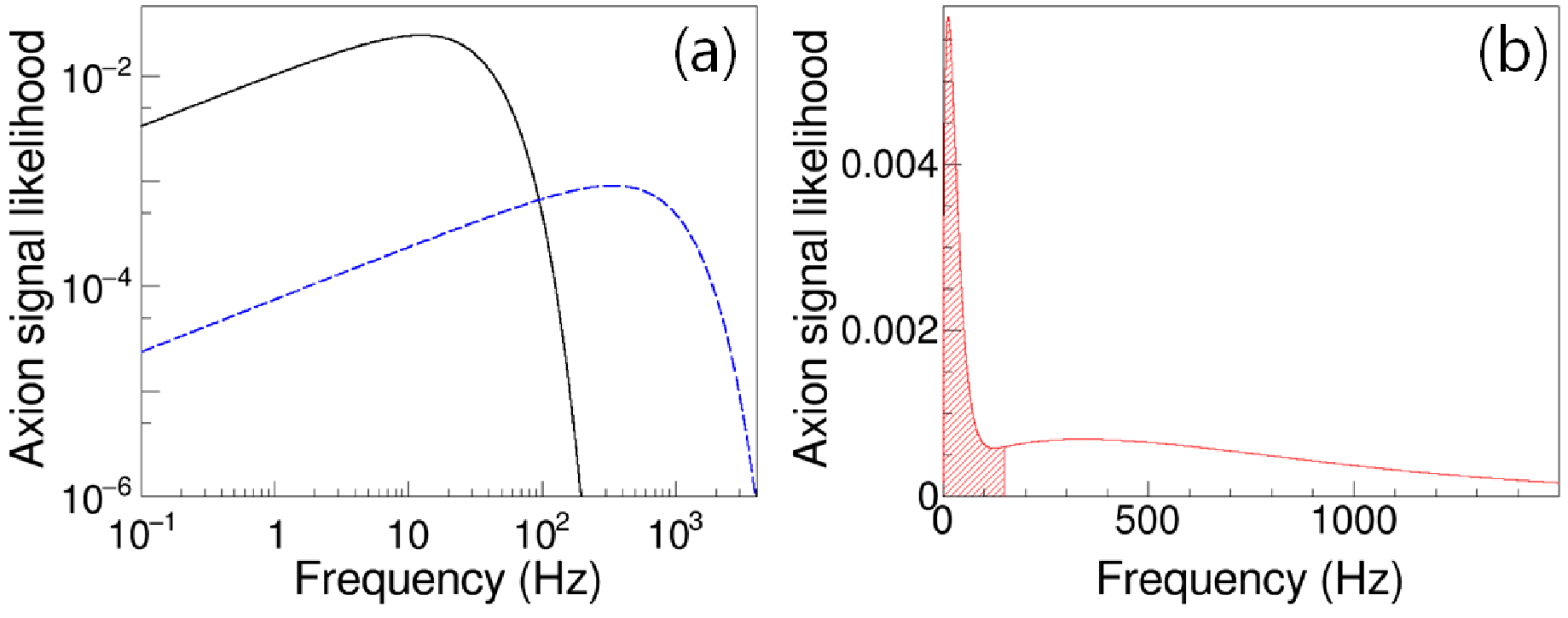}
  \caption{Axion signal shapes for a corresponding frequency of 1.1
    GHz following the SHM (blue dashed line in (a)) and the
    Sagittarius tidal stream model (black solid line in (a)).    
    Red solid line in (b) is the dark matter shape considered in this
    work assuming axion dark matter makes up 100\% of the local dark
    matter density, where the Sagittarius tidal stream model and the
    SHM contribute 23 and 77\%, respectively, and only the red hatched
    region corresponds to the signal model.}  
  \label{FIG:CAPP-12TB-MODEL}
\end{figure}

On the assumption that axions contribute to 100\% of the local dark
matter density and their signal shape follows the blue dashed line in
Fig.~\ref{FIG:CAPP-12TB-MODEL}(a) according to the standard halo model
(SHM)~\cite{AXION_SHAPE}, the CAPP-12TB experiment recently collected
data sensitive to DFSZ axion dark matter, whose mass is around
4.55~$\mu$eV~\cite{12TB_PRL}.
Complementary to the SHM, dark matter of a tidal stream from the
Sagittarius dwarf galaxy would have a velocity $v$ of about 300 km/s
and a velocity dispersion $\delta v$ of about 20 km/s in our Solar
system~\cite{TIDAL_AXION}. Only the 4096-bin search in
Ref.~\cite{ADMX_HR1} corresponds to a search for the Sagittarius tidal
stream of axion dark matter to date.
Another dark matter stream referred to as the ``big flow'' would
possess $v\simeq 480$~km/s and $\delta v\lesssim
53$~m/s~\cite{BIG_FLOW}, and the relevant experimental searches can be
found in Refs.~\cite{ADMX_HR1, ADMX_HR2, ADMX_HR3}.
For dark matter around 4.55~$\mu$eV, the signal widths from the
Sagittarius tidal flow and the big flow would be $\simeq 150$ and
$\lesssim 0.62$~Hz, respectively. The former fits the black solid line
in Fig.~\ref{FIG:CAPP-12TB-MODEL}(a), while the latter is much less
than the frequency resolution bandwidth (RBW) $\Delta f=10$~Hz of the
CAPP-12TB data acquired in March of 2022~\cite{12TB_PRL}.
Considering such signal widths and $\Delta f=10$~Hz, this search seeks
dark matter of the tidal stream~\cite{TIDAL_AXION} around 4.55 $\mu$eV
for the first time, utilizing the same data for our previous
work~\cite{12TB_PRL}. This collateral search without a dedicated
rescan is complementary to our recent SHM search~\cite{12TB_PRL}.

\begin{figure}
  \centering
  \includegraphics[width=0.42\textwidth]{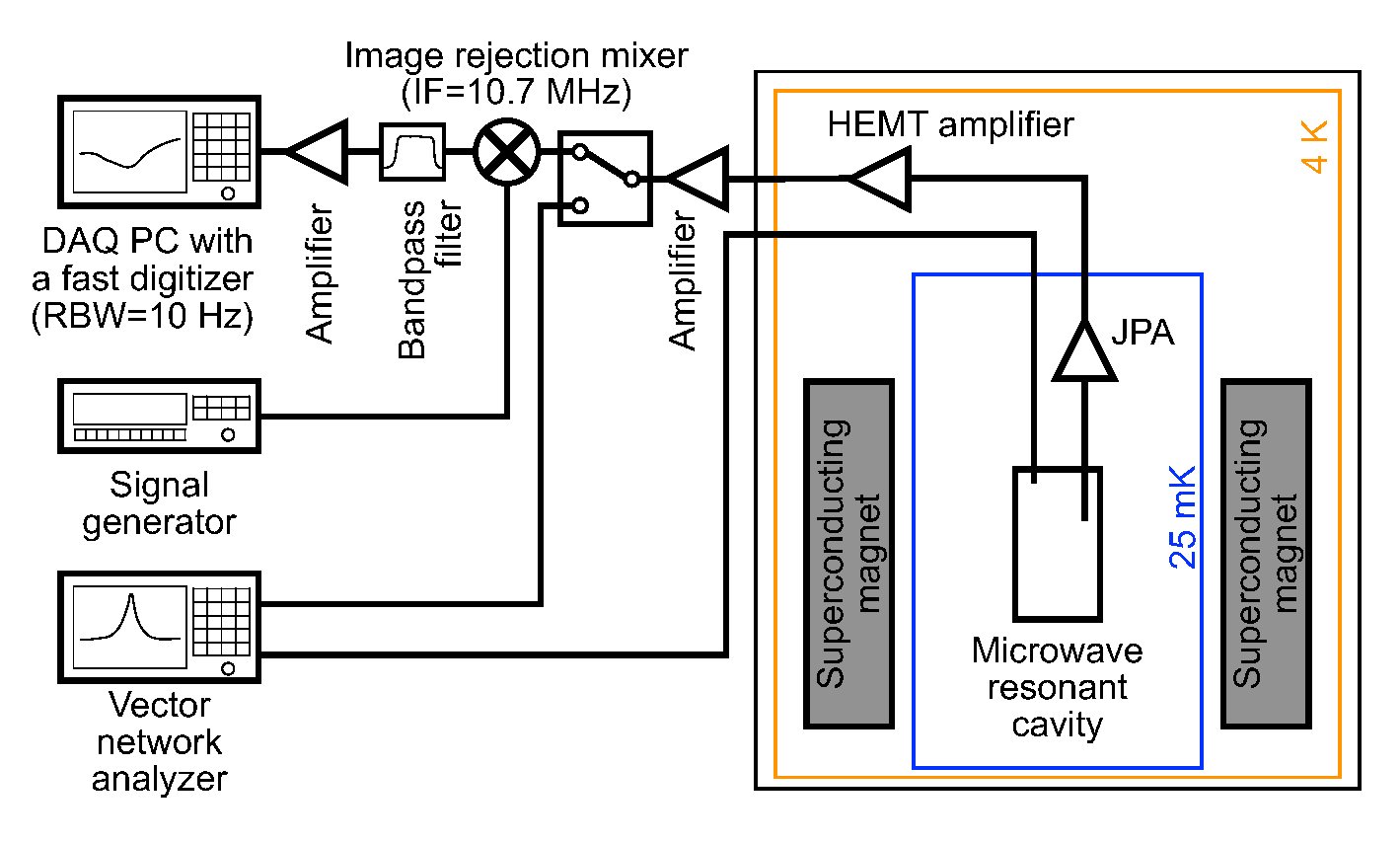}  
  \caption{Schematic of the CAPP-12TB experiment, where HEMT and IF
    stand for High-Electron-Mobility Transistor and Intermediate
    Frequency, respectively.}  
  \label{FIG:CAPP-12TB}
\end{figure}

The experimental parameters of the CAPP-12TB haloscope can be
represented by the expected axion signal power given in
Eq.~(\ref{EQ:PAXION}) and the schematic shown in
Fig.~\ref{FIG:CAPP-12TB}, where Eq.~(\ref{EQ:PAXION}) is valid when
the axion mass $m_a$ matches the frequency of the cavity mode $\nu$
($m_a=h\nu/c^2$) and the cavity mode coupling to the receiver is 2.
\begin{eqnarray}
  P^{a\gamma\gamma}_{a}&=&5.69~{\rm yW}
  \left(\frac{g_\gamma}{0.36}\right)^2
  \left(\frac{B_{\rm rms}}{10.31~{\rm T}}\right)^2 \nonumber \\
  &\times&\left(\frac{V}{36.85~{\rm L}}\right) 
  \left(\frac{C}{0.6}\right) 
  \left(\frac{Q_L}{35~000}\right) \nonumber \\
  &\times&\left(\frac{\nu}{1.1~{\rm GHz}}\right)
  \left(\frac{\rho_a}{0.114~{\rm GeV/cm^3}}\right). 
  \label{EQ:PAXION}
\end{eqnarray}
Equation~(\ref{EQ:PAXION}) also uses the experimental parameters
achieved by the CAPP-12TB experiment which are the rms magnetic field
over a cavity volume $B_{\rm rms}$ of 10.31~T, a cavity volume $V$ of
36.85~L, a cavity-mode-dependent form factor $C$~\cite{EMFF_BRKO} of
0.6, and a loaded quality factor of the cavity mode $Q_L$ of 35~000.
Assuming axion dark matter makes up 100\% of the local dark matter,
where the Sagittarius tidal stream model and the SHM contribute 23 and
77\%~\cite{TIDAL_AXION}, respectively, the detected signal power of
DFSZ axion dark matter of the tidal stream with the SHM contribution
denoted as the red hatched in Fig.~\ref{FIG:CAPP-12TB-MODEL}(b) is
then expected to be 5.69~yW for the axion frequency of~1.1 GHz. This
mixed signal is also similar to that in Ref.~\cite{WIMP_JCAP} and its
power can be obtained by plugging in $\rho_a=0.114$~GeV/cm$^{3}$ in
Eq.~(\ref{EQ:PAXION}), which is the model considered in this work.
Figure~\ref{FIG:CAPP-12TB} not only shows the signal line from the
cavity to the fast data acquisition (DAQ) system~\cite{FDAQ}, but also
another key experimental parameter, the 25~mK physical temperature of
both the cavity and the Josephson Parametric Amplifier (JPA). The fast
digitizer and the signal generator used an external reference
clock~\cite{FS725} to set the frequencies.

The overall operation of the CAPP-12TB experiment and the relevant
measurements can be found in the literature~\cite{12TB_PRL}. Here, we
describe the achieved crucial experimental parameters~\cite{12TB_PRL}
in addition to those shown in Eq.~(\ref{EQ:PAXION}) and
Fig.~\ref{FIG:CAPP-12TB}.
The JPA used by CAPP-12TB~\cite{CAPP-JPA1,CAPP-JPA2} provided gains
and noise temperatures of about 17~dB and 60~mK, respectively, at the
target frequencies over the search range, where the target frequency
is the central frequency of each individual power spectrum and $\nu$
was tuned to it.
The gains and noise temperatures of the receiver chain other than the
JPA were about 104~dB and 1.2~K, respectively. The total gains of the
receiver chain were close to 121~dB around $\nu$.
The total gain in the measured power was removed to obtain the
total system noise temperature $T_n$ from the cavity and the receiver
chain.
The gain corrected power of each power spectrum was then parameterized
using a Savitzky-Golay (SG) filter~\cite{SGF}, and the extracted $T_n$
around $\nu$ was approximately 215~mK as a roughly Lorentzian
peak~\cite{12TB_PRL}.

For each frequency step, we took 40 power spectra, with each being the
average of 128 individual spectra~\cite{FDAQ}.
Individual spectra with an analysis frequency span of 150~kHz were
measured over a 0.1~s interval, hence, $\Delta f=10$~Hz, as mentioned
above, and the integrated time $\Delta t$ for the measurement of the
power at each frequency $P_{f_i}$ was 512~s. During this DAQ duration
of 512~s, $\nu$ drifts with a spread of about 290~Hz due to our
experimental imperfections, while the axion
signal shape is almost invariant under the Earth motions as explained
below. As our cavity bandwidth is about 31~kHz for $\nu$ around 1.1
GHz, the axion signal power at $\nu$ and that at 290~Hz away from
$\nu$ differs by less than 0.1\% in the case of $\nu$ drift, which is
negligible to our result.

The 150~kHz analysis span and our frequency tuning steps of 10~kHz
create 15 power spectra overlaps in most of the frequency
range, meaning that the total $\Delta t$ for
the $P_{f_i}$ measurement is at least
7680~s. During the $P_{f_i}$ measurement time of 10~000~s considering
the cavity tuning and the relevant measurements conservatively, the
rotational and orbital motions of Earth with corresponding speeds of
0.4 and 30~km/s can shift the signal frequency at most by about 1.07
and 0.22~Hz, respectively. Hence, the signal broadening caused by the
motions of Earth are negligible compared to the RBW of 10~Hz and to
the expected signal width of about 150~Hz.
\begin{figure*}
  \centering
  \subfigure{\includegraphics[width=0.24\textwidth]{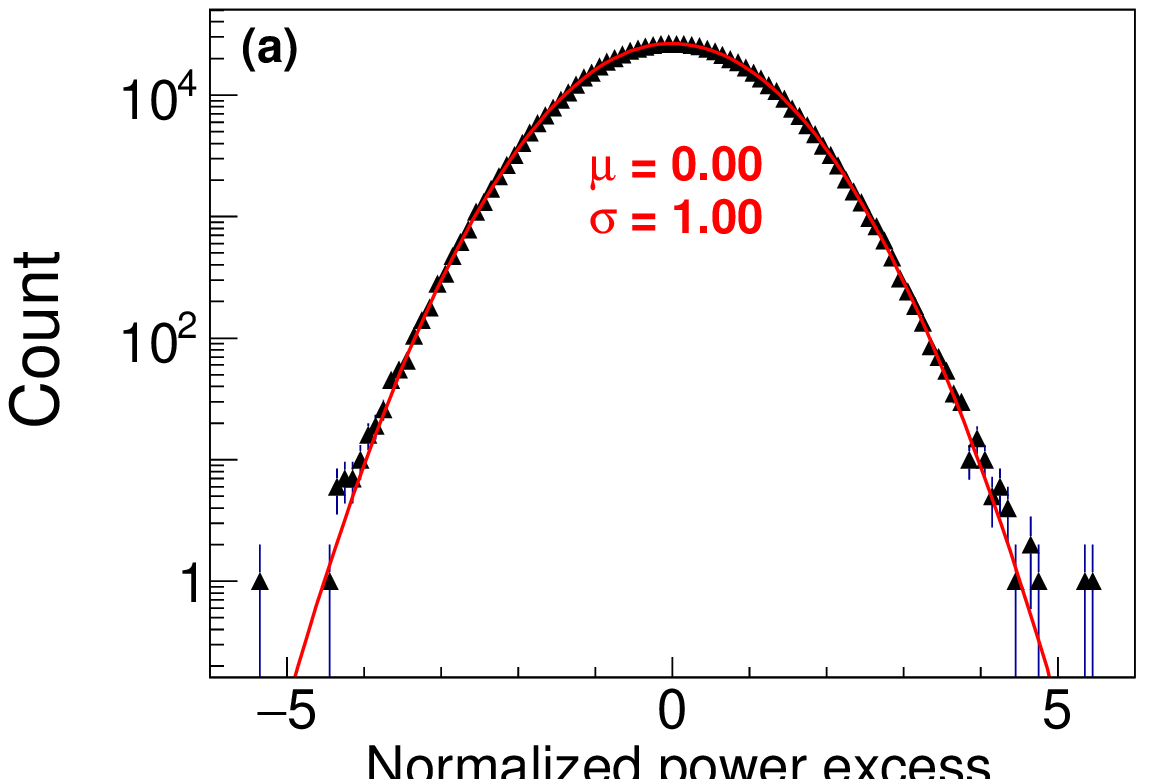}}
  \subfigure{\includegraphics[width=0.24\textwidth]{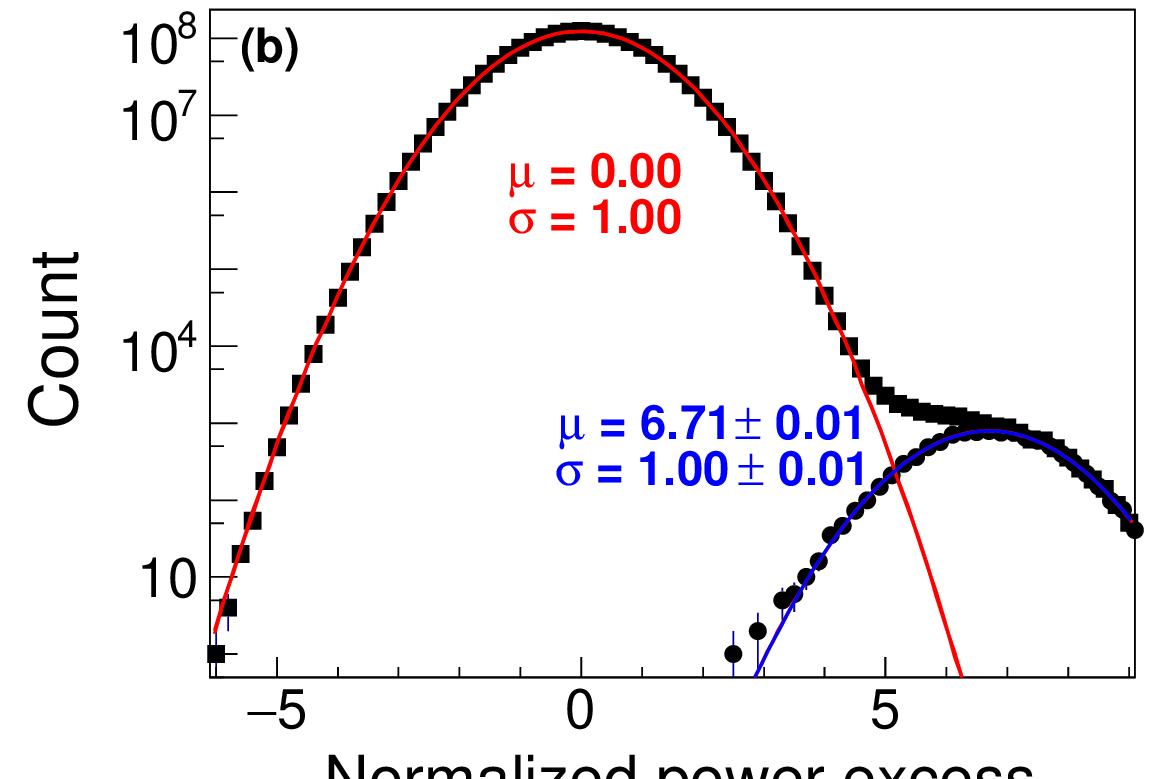}}
  \subfigure{\includegraphics[width=0.24\textwidth]{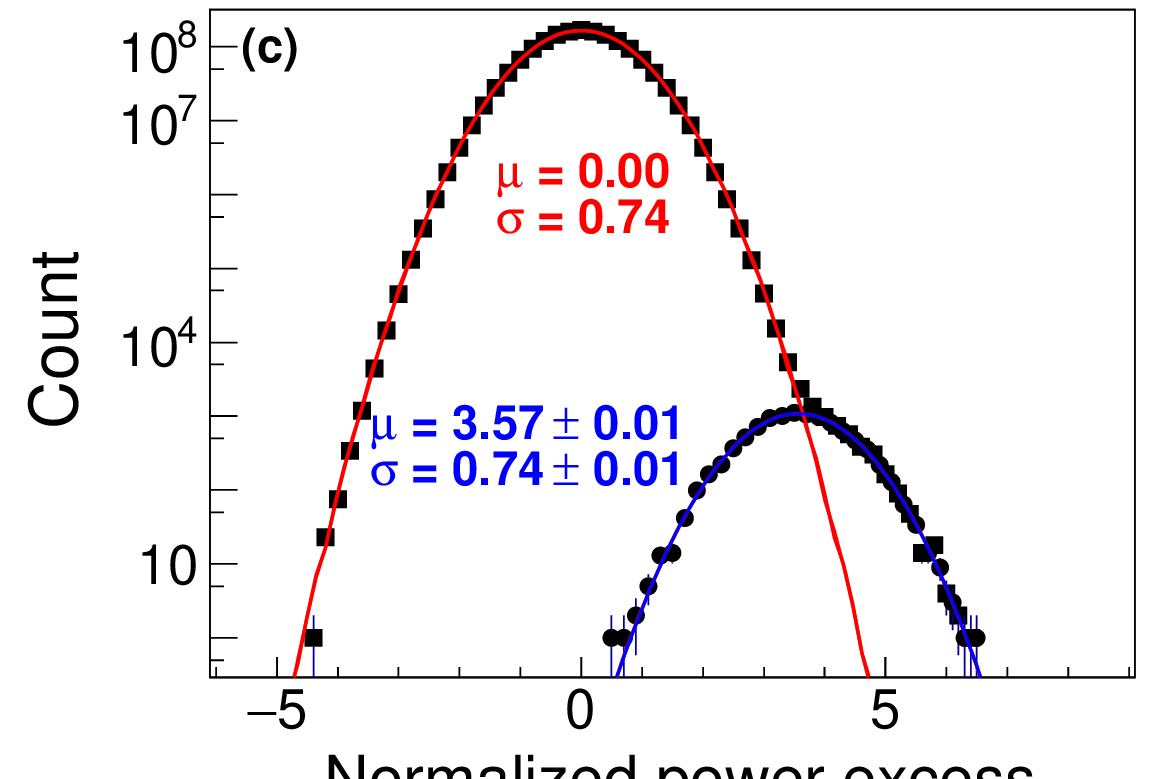}}
  \subfigure{\includegraphics[width=0.24\textwidth]{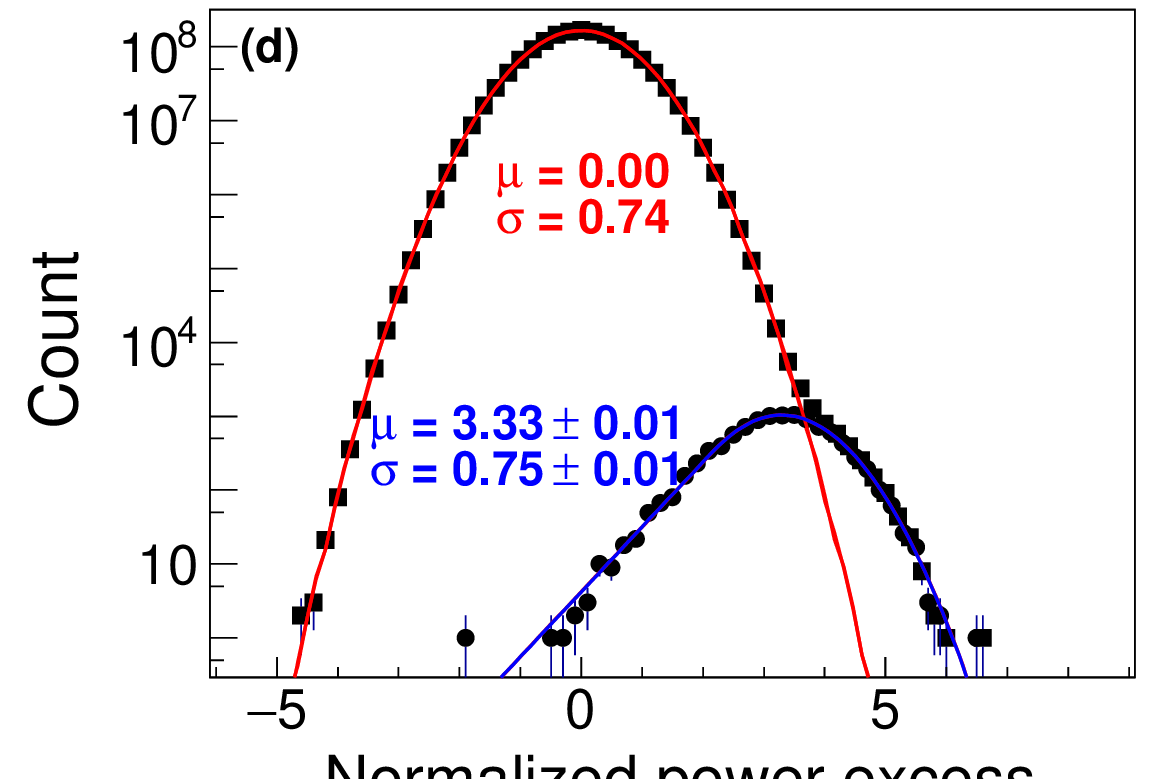}}
  \caption{Triangles in (a) show the normalized power excess
    distribution of the normalized grand power spectrum from the
    CAPP-12TB data after applying a frequency-independent scale
    factor of 0.74.        
    (b), (c), and (d) show the normalized power excess distributions
    of 10~000 normalized grand power spectra from the 10~000 simulated
    CAPP-12TB experiments.    
    Rectangles in (b), (c), and (d) are from all frequencies of
    the normalized grand spectra, while the circles there are from the
    frequency with the simulated axion signals.
    (b) and (c) were obtained without a filtering procedure, while
    (d) was obtained with filtering.    
    (b) was obtained after background subtraction with a perfect
    fit, while (c) and (d) were obtained with our SG filter with $d=4$
    and $W=37$. Lines are a Gaussian fit resulting in $\mu$ (mean) and
    $\sigma$ (width), except for the Crystal Ball line
    shape~\cite{CBLine} fit to the circle distribution in (d).}  
  \label{FIG:CAPP-12TB-PULLS}
\end{figure*}

Given that the signal frequency shifts are much less than the RBW of
10~Hz for the duration of the $P_{f_i}$ measurement, $P_{f_i}$ can be
calculated as the weighted average of the powers measured from all
spectra having that frequency, which allows us to use, without a
further RBW reduction process, the similar analysis procedure for the
axion dark matter search of the
SHM~\cite{ADMX_ANAL, HAYSTAC_ANAL, CAPP_ANAL}.
Owing to the absence of a dedicated rescan schedule, our analysis
strategy here seeks to let the most significant power excess have
significance under a certain threshold value and as small as possible.
Having a narrower signal window and lower $\rho_a$, the
signal-to-noise ratio (SNR) is na\"{\i}vely expected to be consistent
approximately with that in our previous work~\cite{12TB_PRL}.
If a threshold of 3.718 of the normalized power excess to get a
one-side 90\% upper limit corresponding to the expected SNR of 5 would
be applicable without rescanning, we could expect a search sensitive
to the Sagittarius tidal stream of DFSZ axion dark matter considered
in this study at a 90\% confidence level (CL).

First, we applied a similar filtering procedure in
Ref.~\cite{HAYSTAC_ANAL} to remove narrow spikes in each power
spectrum.
Each power spectrum was parameterized via SG smoothing with
a polynomial of degree 4 ($d=4$) in a 75-point window ($2W+1=75$) at
$\Delta f=10$~Hz for both the filtering (see the details in
Appendix~\ref{FILTERING}) and background subtraction thereafter, where
the 75-point window corresponds to five times the signal window used
for this search.
We found that the normalized power excess obtained right after
background subtraction from each power spectrum follows a Gaussian
distribution whose width is not 1, but 0.97.
This narrower Gaussian width resulted from the SG filter parameters of
$d=4$ and $W=37$ used in this work, because we found no such bias for
SG parameters of $d=4$ and $W=1000$ used in our previous
work~\cite{12TB_PRL}.
Our simulation data explained below also predict the bias induced
from the SG filter with $d=4$ and $W=37$, and thus a possible
systematic effect is discussed there as well.
After filtering and background subtraction, all of the power spectra
were combined as a single power spectrum. We then summed three
nonoverlapping spectral lines so that $\Delta f=30$~Hz, and this
spectrum is referred to as the ``combined power spectrum'' (see the
details in Appendix~\ref{VCOMB}).
This RBW reduction to 30~Hz ensures a smaller threshold value because
it decreases the significance of power excess, albeit it was a
compromise with the decrease in SNR that accompanies this
process~\cite{12TB_PRL}.
Our ``grand power spectrum'' was constructed from the combined power
spectrum by co-adding~\cite{ADMX_ANAL} five continuously adjacent
30~Hz power spectral lines. Each spectral line was weighted by the
signal shape~\cite{HAYSTAC_ANAL}, which is indicated by the red
hatched region in Fig.~\ref{FIG:CAPP-12TB-MODEL}(b).
Figure~\ref{FIG:CAPP-12TB-PULLS}(a) shows the normalized power excess
distribution from the grand power spectrum following the standard
Gaussian after correcting for a frequency-independent scale factor of
0.74 to remove the known bias in power excess induced by background
subtraction~\cite{CAPP_ANAL}.
From such a standard Gaussian distribution, we found that the greatest
significance of the power excess is less than 5.4 and thus were able
to exclude all of the power excess by applying a threshold of 5.4
without rescanning.
For the most significant power excess, we calculated the $\chi^2$
probability to check the signal compatibility (see the details in
Appendix~\ref{SIGNAL_TEST}). From it, we concluded that the excess is
not compatible with the signal model in this work and thus no dark
matter of the tidal stream was found around 4.55~$\mu$eV.

We estimated the SNR efficiency and other possible systematic effects
with 10~000 simulated CAPP-12TB experiments with simulated axion
signals at an arbitrary frequency within the search range. The input
signal power and shape of the simulated axion signals follow
Eq.~(\ref{EQ:PAXION}) and the red hatched region in
Fig.~\ref{FIG:CAPP-12TB-MODEL}(b), respectively.
In order to avoid the aforementioned bias observed in the power
excess immediately after background subtraction by the SG filter above,
we instead used backgrounds parameterized by another SG filter with
$d=4$ and $W=1000$ as our simulation inputs.
Figures~\ref{FIG:CAPP-12TB-PULLS}(b) and (c) show the distributions of
the normalized power excess from 10~000 grand power spectra from the
10~000 simulated CAPP-12TB experiments without the filtering
procedure, where the former was obtained by background subtraction
using the simulation inputs, i.e., a perfect fit, and the latter was
obtained using the SG filter with $d=4$ and $W=37$. The rectangles in
Figs.~\ref{FIG:CAPP-12TB-PULLS}(c) and (d) show a width of 0.74,
agreeing with the CAPP-12TB data.
The circles in Figs.~\ref{FIG:CAPP-12TB-PULLS}(b) and (c) are from the
frequency with the simulated axion signals. From these, we estimated
that our background subtraction efficiency was about 72\%.
Figure~\ref{FIG:CAPP-12TB-PULLS}(d) is identical to
Fig.~\ref{FIG:CAPP-12TB-PULLS}(c), but with the filtering procedure
applied. The filtering efficiency was extracted and determined to be
about 92\% from the distributions of the circles in
Figs.~\ref{FIG:CAPP-12TB-PULLS}(c) and (d).
A fast Fourier transform (FFT) returns discrete power lines only at
the set frequencies and thus can be inefficient unless the input
signal frequencies match those frequencies. The efficiency depends on
the number of spectral lines with the signal power, e.g., it is at
worst 40.5\% for a single spectral line search~\cite{ADMX_HR1}.
For the search here with 15 spectral lines, the minimum efficiency was
estimated to be about 97\% for the 5~Hz offset between the input and
the set frequencies, which included at most 0.1\% inefficiency due to
the RBW reduction from 10 to 30~Hz.
Our total SNR efficiency is approximately 64\% taking into account the
three sources, 72\% from background subtraction, 92\% from narrow
spike filtering, and 97\% from FFT with maximum signal misalignment.
Further frequency-dependent SNR degradation due to the additional line
attenuation insensitive to our noise calibration~\cite{12TB_PRL} was
also reflected in our exclusion limits.
\begin{figure}
  \centering
  \includegraphics[width=0.48\textwidth]{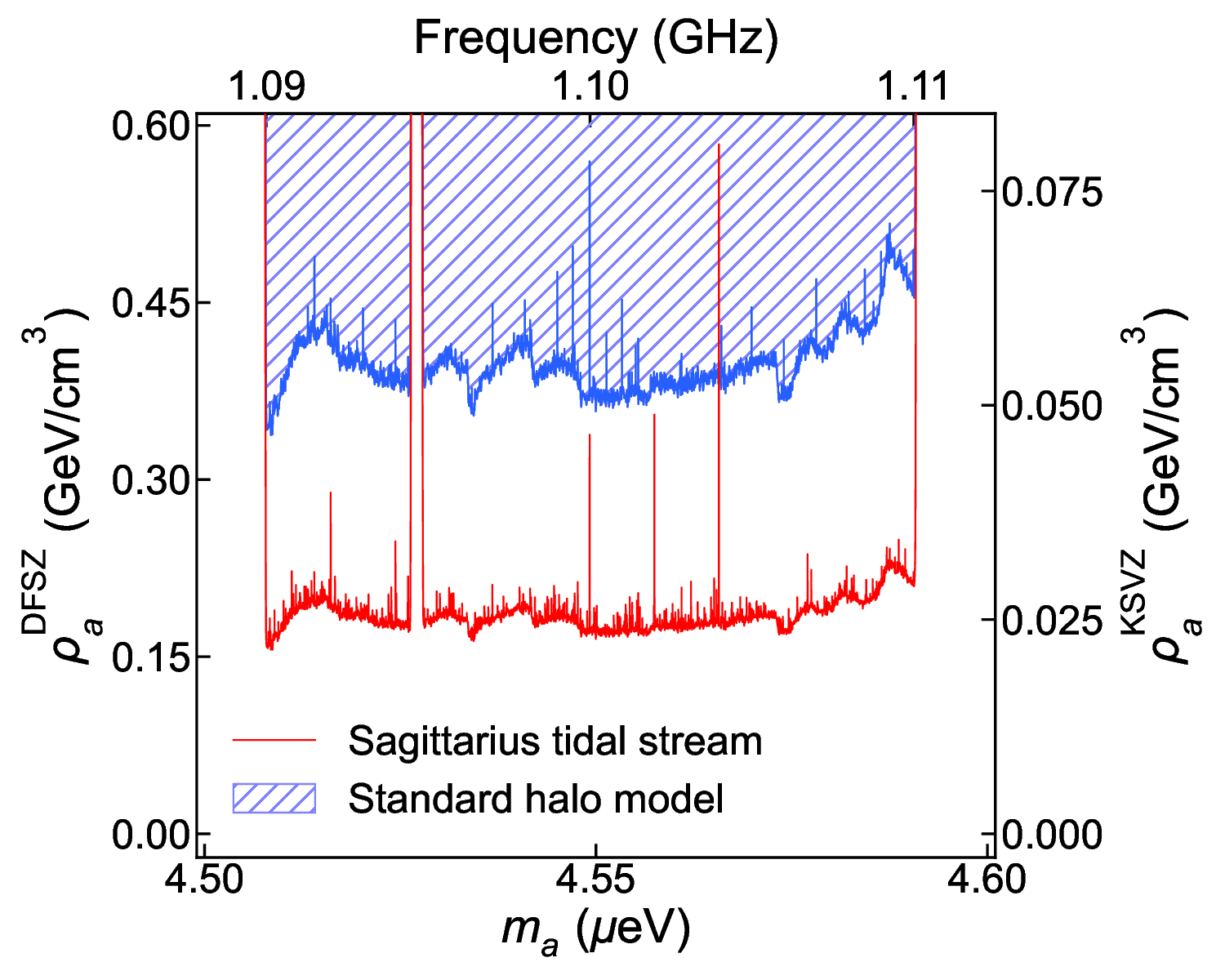}  
  \caption{Assuming axion dark matter makes up 100\% of the local dark
    matter density, the blue hatched region shows the exclusion limits
    for the axion dark matter densities, $\rho^{\rm DFSZ}_a$ in left
    and $\rho^{\rm KSVZ}_a$ in right, at a 90\% CL~\cite{12TB_PRL} and
    the red solid line shows those achieved by this work.
    The former used the SHM~\cite{AXION_SHAPE} contribution of 100\%
    (blue dashed line in Fig.~\ref{FIG:CAPP-12TB-MODEL}(a)) and the
    latter the Sagittarius tidal stream model and the SHM
    contributions of 23 and 77\%, respectively~\cite{TIDAL_AXION}, but
    only considering the red hatched region in
    Fig.~\ref{FIG:CAPP-12TB-MODEL}(b).    
    No results are available around an axion mass of 4.527~$\mu$eV due
    to mode crossing.    
    The spikes are less sensitive frequency points with fewer
    statistics resulting from the filtering procedure mentioned in the
    text.}  
  \label{FIG:CAPP-12TB-LIMIT}
\end{figure}

The 10~000 simulated CAPP-12TB experiments were also used to estimate
a possible systematic effect that can come from the bias observed in
the power excess right after background subtraction using our SG
filter with $d=4$ and $W=37$. Using the large-statistic simulation
data, we were able to separate the systematic effects from the
statistical fluctuations by comparing the results with the perfect fit
and our SG filter. The noise fluctuations resulting from the perfect
fit follow the radiometer equation~\cite{DICKE}, whereas we found that
those resulting from the SG filter do not.
The additional contribution to the noise fluctuations resulting from
the SG filter was extracted and found to be about 0.5\% of the
statistical fluctuations.
The systematic uncertainty of 6\% in the noise temperature measurement
was dominant~\cite{12TB_PRL} and this factor was taken into account in
our exclusion limits.

For $4.51<m_a<4.59$ $\mu$eV, we set the 90\% upper limits of the
densities of axion dark matter of the Sagittarius tidal
stream~\cite{TIDAL_AXION}, as shown in
Fig.~\ref{FIG:CAPP-12TB-LIMIT}. Dark matter of the tidal streams was
ruled out for densities of $\rho_a\gtrsim0.184$ and
$\gtrsim0.025$~GeV/cm$^3$ for DFSZ and KSVZ axions, respectively, at a
90\% CL over the search range. The exclusion limits for DFSZ axions
are less sensitive than the aforementioned expected value of
0.114~GeV/cm$^3$, which resulted from the lower SNR efficiency and the
larger threshold compared to our previous axion dark matter search of
the SHM~\cite{12TB_PRL}.
This was an inevitable side-effect of reprocessing the data from
Ref.~\cite{12TB_PRL} without additional scanning time. Nevertheless,
our result excluded the Sagittarius tidal stream of KSVZ axion dark
matter for the first time, down to about 5.6\% of the local dark
matter density assuming the dark matter model considered in this
work~\cite{TIDAL_AXION}.

In summary, we report the first search for the Sagittarius tidal
stream of axion dark matter around 4.55 $\mu$eV using CAPP-12TB
haloscope data acquired in March of 2022.
We excluded the Sagittarius tidal stream of DFSZ and KSVZ axion dark
matter densities of $\rho^{\rm DFSZ}_a\gtrsim0.184$~GeV/cm$^{3}$ and
$\rho^{\rm KSVZ}_a\gtrsim0.025$~GeV/cm$^{3}$, respectively, over a mass
range from 4.51 to 4.59~$\mu$eV at a 90\% CL.

This work was supported by the Institute for Basic Science (IBS) under
Project Code No. IBS-R017-D1-2023-a00 and Japan Science and Technology
Agency ERATO (Grant No. JPMJER {\bf 1601}). A. F. van Loo was
supported by a JSPS postdoctoral fellowship.

\appendix
\section{Narrow spike filtering}\label{FILTERING}
\begin{figure*}
  \centering
  \includegraphics[width=1.0\textwidth]{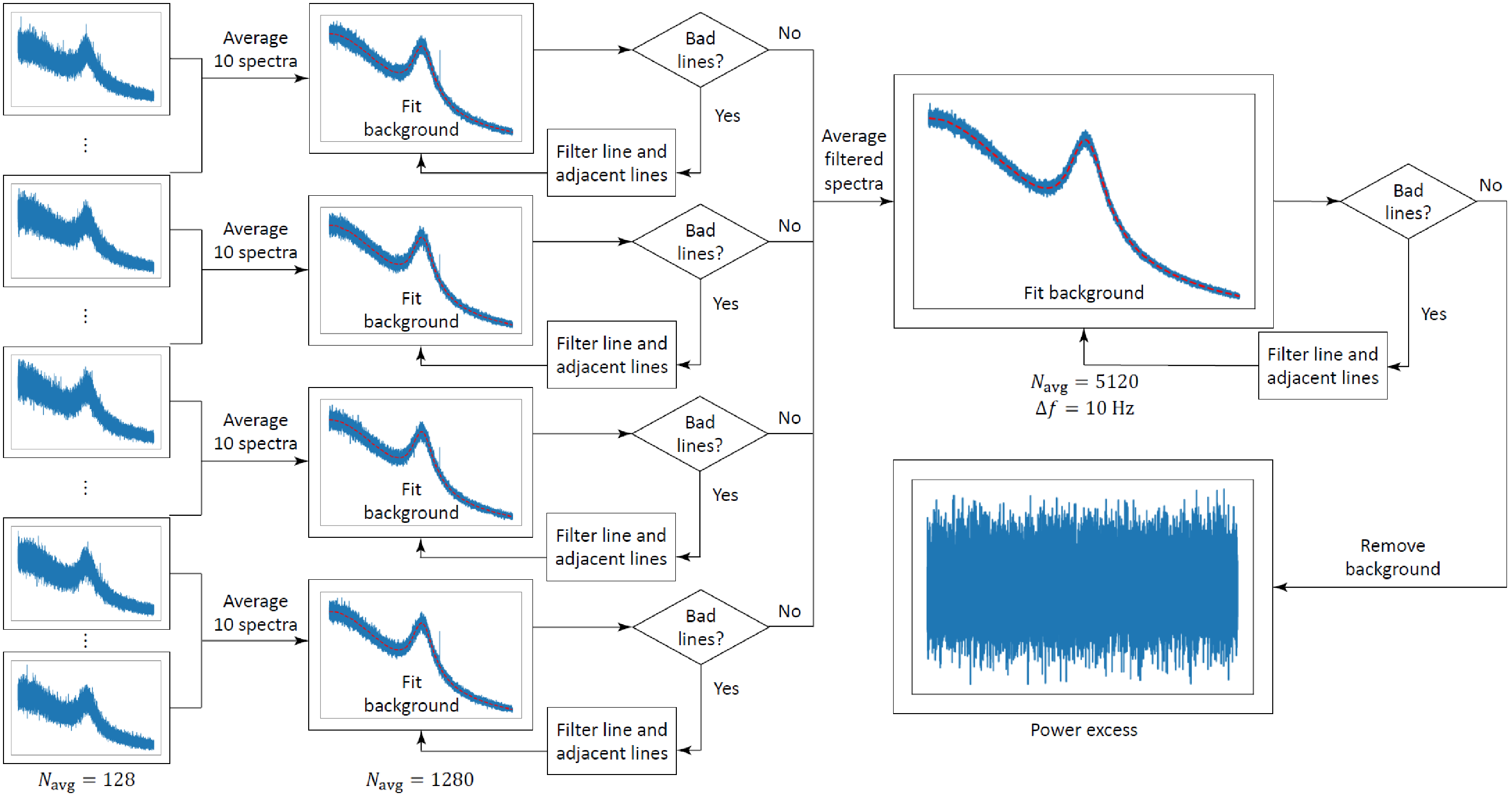}
  \caption{Procedure of the narrow spike filtering. Red dashed lines
    show the background fits by our SG filter with $d=4$ and $W=37$.}  
  \label{FIG:FILTERING}
\end{figure*}

As depicted in Fig.~\ref{FIG:FILTERING}, the 40 power spectra taken at
each tuning step were sampled into four groups and the ten power
spectra in each group were averaged to result in four power spectra so
that narrow spikes which may not show up in a single spectrum would do
so after ten averages. In order to obtain a threshold value as small
as possible, we applied tighter filtering conditions.
Each of the four power spectra was fitted using the SG filter with
$d=4$ and $W=37$ and the normalized power excesses over 3.5 were then
removed in each of the four power spectra.
A neighboring frequency point on both sides of the narrow spike
passing the threshold 3.5 was also eliminated. After averaging the
four power spectra, filtering was applied again to the averaged power
spectrum using the same conditions described above.
\section{Combined power spectrum}\label{VCOMB}
\begin{figure}
  \centering
  \includegraphics[width=0.41\textwidth]{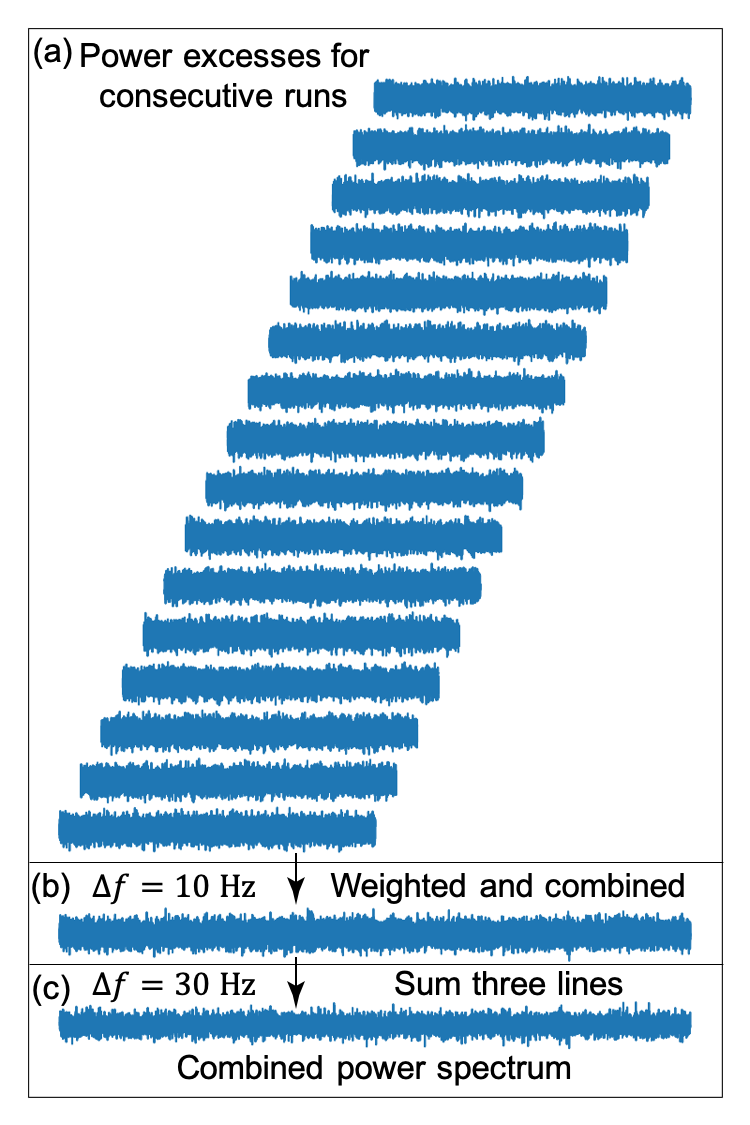}
  \caption{Construction of the combined power spectrum. (a) shows the
    background-subtracted power spectra shown in
    Fig.~\ref{FIG:FILTERING} for consecutive frequency steps. (b) and
    (c) are the combined power spectrum with $\Delta f$ of 10 and
    30~Hz, respectively, where the excess at each frequency was
    obtained as the weighted average of the excess from all the
    relevant individual spectra shown in (a).}  
  \label{FIG:V_COMBI}  
\end{figure}

Figure~\ref{FIG:V_COMBI} illustrates the construction of the combined
power spectrum. Each of the consecutive background-subtracted power
spectra corresponds to a power excess shown in
Fig.~\ref{FIG:FILTERING}. All of them and the associated errors were
rescaled by the expected total axion signal power, where the rescaling
across the spectrum follows the cavity line shape reflecting the
corresponding $Q_L$~\cite{12TB_PRL, HAYSTAC_ANAL}. The excess at each
frequency was then obtained as the weighted average of the excess from
all the relevant individual spectra that are combined.
\section{Signal compatibility test}\label{SIGNAL_TEST}
\begin{figure}
  \centering
  \includegraphics[width=0.43\textwidth]{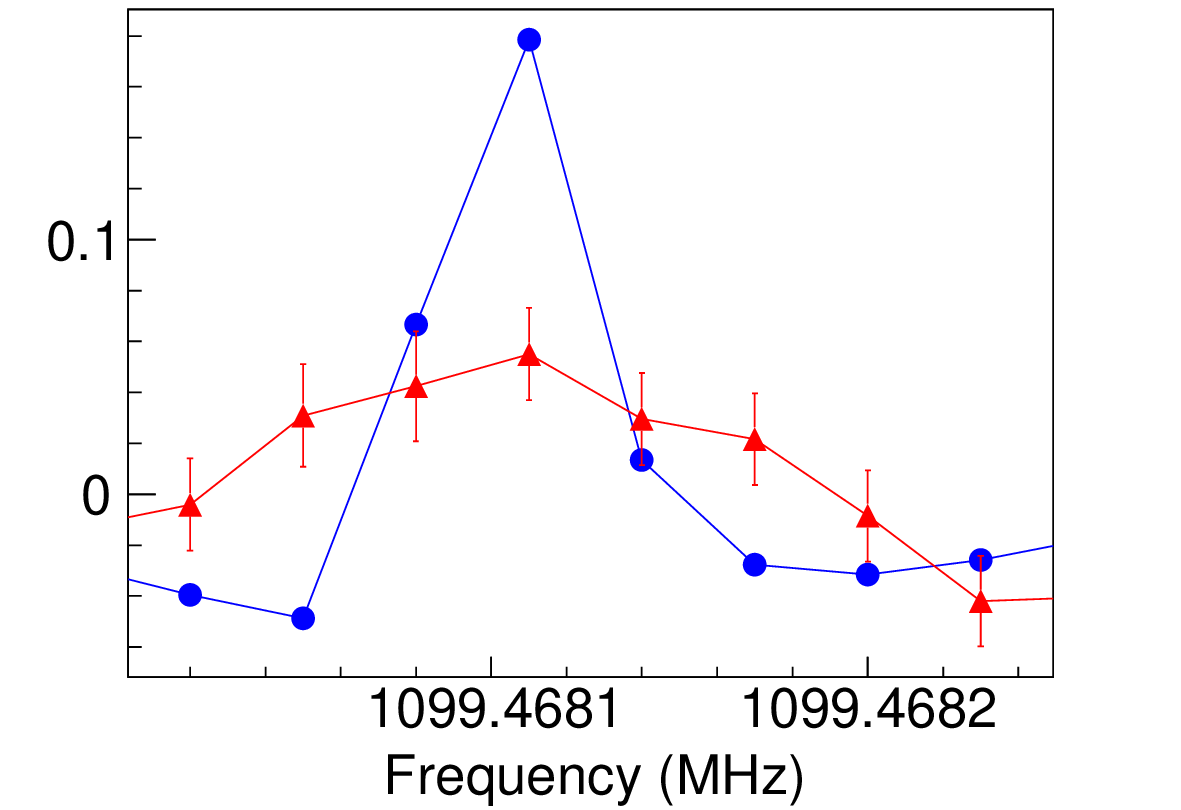}
  \caption{Red solid triangles with error bars are data from the
    combined power spectrum and blue solid circles the simulation
    expectations. The signal compatibility was checked with five
    continuously neighboring excesses from the 3rd to the 7th
    excess corresponding to the signal region. Note that the excesses
    and errors are dimensionless due to the rescaling mentioned in
    Appendix~\ref{VCOMB} and the negative excesses can be found due to
    background subtraction in the presence of axion signal, as also
    seen in Ref.~\cite{CAPP_ANAL}.}  
  \label{FIG:SIG_COMP}
\end{figure}

We checked the signal compatibility by comparing the two signal
shapes, one is from data and the other from the model considered in
this work. Data points are red solid triangles with error bars in
Fig.~\ref{FIG:SIG_COMP} from the combined power spectrum which retains
the signal line shape.
The simulation expectations including our signal model reflecting the
significance of the excess observed from data are blue solid circles
in Fig.~\ref{FIG:SIG_COMP}.
Our $\chi^2=\sum_i\frac{(d_i-e_i)^2}{\sigma^2_{d_i}}$, where $d_i$ and
$e_i$ are data and the expectations, respectively, and $\sigma_{d_i}$
the $d_i$ errors.
Note that the $\sigma_{d_i}$ are the rescaled errors after the
background estimation from the fit shown as the red dashed line in
Fig.~\ref{FIG:FILTERING}, while the $d_i$ and $e_i$ are the rescaled
power excesses after background subtraction by the fit.
The $\chi^2$ was built with five continuously neighboring excesses,
where the frequency with the greatest significance was aligned with
the 2nd of the five excesses according to the signal shape, as denoted
by the red hatched region in Fig.~\ref{FIG:CAPP-12TB-MODEL}(b). The
calculated $\chi^2$ probability with a degree of freedom of 4 was
about $1.8\times10^{-9}$, which implies the two signal shapes are not
compatible with each other.


\begin{thebibliography}{99}

\bibitem{PLANCK}
  P. A. R. Ade {\it et al.} (Planck Collaboration), Astron. Astrophys. \textbf{594}, A13 (2016). 

\bibitem{CDM-EVIDENCE}
  V. Rubin and W. K. Ford Jr., ApJ \textbf{159}, 379 (1970);
  Douglas Clowe {\it et al.}, ApJ \textbf{648}, L109 (2006).

\bibitem{AXION}
  S. Weinberg, Phys. Rev. Lett. \textbf{40}, 223 (1978);
  F. Wilczek, Phys. Rev. Lett. \textbf{40}, 279 (1978).

\bibitem{PQ}
  R. D. Peccei and H. R. Quinn, Phys. Rev. Lett. \textbf{38}, 1440 (1977).

\bibitem{strongCP}
  G. 't Hooft, Phys. Rev. Lett, {\bf 37}, 8 (1976); 
  Phys. Rev. D {\bf 14}, 3432 (1976); {\bf 18}, 2199(E) (1978); 
  J. H. Smith, E. M. Purcell, and N. F. Ramsey, Phys. Rev. \textbf{108}, 120 (1957);
  W. B. Dress, P. D. Miller, J. M. Pendlebury, P. Perrin, and N. F. Ramsey, Phys. Rev. D {\bf 15}, 9 (1977); 
  I. S. Altarev {\it et al.}, Nucl. Phys. \textbf{A341}, 269 (1980). 

\bibitem{CDM_LOW}
  J. Preskill, M. B. Wise, and F. Wilczek, Phys. Lett. \textbf{120B}, 127 (1983);
  L. F. Abbott and P. Sikivie, Phys. Lett. \textbf{120B}, 133 (1983);
  M. Dine and W. Fischler, Phys. Lett. \textbf{120B}, 137 (1983).
  
\bibitem{sikivie}
  P. Sikivie, Phys. Rev. Lett. \textbf{51}, 1415 (1983); Phys. Rev. D {\bf 32}, 2988 (1985).

\bibitem{KSVZ}
  J. E. Kim, Phys. Rev. Lett. \textbf{43}, 103 (1979);
  M. A. Shifman, A. I. Vainshtein, and V. I. Zakharov, Nucl. Phys. \textbf{B166}, 493 (1980). 

\bibitem{DFSZ}
  A. R. Zhitnitskii, Yad. Fiz. {\bf 31}, 497 (1980) [Sov. J. Nucl. Phys. \textbf{31}, 260 (1980)];
  M. Dine, W. Fischler, and M. Srednicki, Phys. Lett. \textbf{104B}, 199 (1981).

\bibitem{AXION_SHAPE}
  Michael S. Turner, Phys. Rev. D \textbf{42}, 3572 (1990).  

\bibitem{12TB_PRL}
  Andrew K. Yi {\it et al.}, Phys. Rev. Lett. \textbf{130}, 071002 (2023).  

\bibitem{TIDAL_AXION}
  K. Freese, P. Gondolo, H. J. Newberg, and M. Lewis, Phys. Rev. Lett. \textbf{92}, 111301 (2004);
  K. Freese, P. Gondolo, and H. J. Newberg, Phys. Rev. D \textbf{71}, 043516 (2005).  

\bibitem{ADMX_HR1}
  L. Duffy {\it et al.}, Phys. Rev. Lett. \textbf{95}, 091304 (2005);
  Phys. Rev. D \textbf{74}, 012006 (2006).

\bibitem{BIG_FLOW}
  P. Sikivie, Phys. Lett. B \textbf{567}, 1 (2003).

\bibitem{ADMX_HR2}
  J. Hoskins {\it et al.}, Phys. Rev. D \textbf{84}, 121302(R) (2011).

\bibitem{ADMX_HR3}
  J. Hoskins {\it et al.}, Phys. Rev. D \textbf{94}, 082001 (2016).
  
\bibitem{EMFF_BRKO}
  B. R. Ko {\it et al.}, Phys. Rev. D \textbf{94}, 111702(R) (2016).  

\bibitem{WIMP_JCAP}
  Chris W. Purcell, Andrew R. Zentner, and Mei-Yu Wang, JCAP \textbf{08} (2012) 027.

\bibitem{FDAQ}
  S. Ahn {\it et al.}, J. Instrum. \textbf{17}, P05025 (2022).

\bibitem{FS725}
  FS725 Rubidium Frequency Standard, Stanford Research Systems,
  1290-D Reamwood Avenue, Sunnyvale, CA 94089, USA.

\bibitem{CAPP-JPA1}
  T. Yamamoto {\it et al.}, Appl. Phys. Lett. \textbf{93}, 042510 (2008).

\bibitem{CAPP-JPA2}
  \c{C}a\u{g}lar Kutlu {\it et al.}, Supercond. Sci. Technol. \textbf{34}, 085013 (2021).

\bibitem{SGF}
  A. Savitzky and M. J. E. Golay, Anal. Chem. \textbf{36}, 1627 (1964).

\bibitem{ADMX_ANAL}
  S. J. Asztalos {\it et al.}, Phys. Rev. D \textbf{64}, 092003 (2001).

\bibitem{HAYSTAC_ANAL}
  B. M. Brubaker, L. Zhong, S. K. Lamoreaux, K. W. Lehnert, and K. A. van Bibber, Phys. Rev D \textbf{96}, 123008 (2017).

\bibitem{CAPP_ANAL}
  S. Ahn, S. Lee,  J. Choi, B. R. Ko, and Y. K. Semertzidis, J. High Energy Phys. \textbf{04} (2021) 297.
  
\bibitem{CBLine}
  T. Skwarnicki, Ph.D. thesis, Cracow, INP, 1986.
  
\bibitem{DICKE}
  R. H. Dicke, Rev. Sci. Instrum. \textbf{17}, 268 (1946).

\end{thebibliography}
\end{document}